\documentclass[aps,prd,reprint,nofootinbib,superscriptaddress]{revtex4-2}

\usepackage[T1]{fontenc}
\usepackage[utf8]{inputenc}
\usepackage{amsmath,amssymb,amsthm,bm,mathtools}
\usepackage{graphicx}
\usepackage{xcolor}
\usepackage{booktabs}
\usepackage{hyperref}
\hypersetup{hidelinks}
\usepackage{placeins}

\newcommand{\dd}{\mathrm{d}}
\newcommand{\ee}{\mathrm{e}}
\newcommand{\ech}{e_{\mathrm{ch}}}
\newcommand{\Order}{\mathcal{O}}
\newcommand{\kb}{k_{\mathrm B}}
\newcommand{\lp}{\ell_{\mathrm P}}
\newcommand{\BH}{BH}
\newcommand{\BHs}{BHs}

\begin{document}

\title{Classical corrections to \BH{} entropy II}

\author{Daulet Berkimbayev}
\email{daulet9432@gmail.com}
\affiliation{Institute of Physics, Al-Farabi Kazakh National University, Al-Farabi Avenue 71, Almaty 050040, Kazakhstan}

\author{Martin Blaschke}
\email{martin.blaschke@physics.slu.cz}
\affiliation{Institute of Physics and Research Centre of Theoretical Physics and Astrophysics, Silesian University in Opava, Bezru\v{c}ovo n\'am\v{e}st\'i~13, CZ-746\,01 Opava, Czech Republic}

\begin{abstract}
We reconsider the classical one bit absorption model of black hole growth as a discrete recursion rather than a continuum equation. Here ``one bit'' denotes one elementary absorption unit in natural logarithmic units. For a Schwarzschild--Tangherlini \BH{}, the discrete treatment yields the expected area scaling together with a logarithmic correction whose coefficient is fixed by the dimensional dependence of the mass step. We then extend the construction to the Reissner--Nordstr\"om case at fixed charge and show that the logarithmic coefficient acquires an explicit charge dependence. The fixed charge sector can be read as the asymptotic neutral growth stage after a charged seed has already been formed, so realistic charged absorption affects the initial cutoff rather than the large mass recursion for weakly charged final states. Ordered charged, oppositely charged, and neutral labels define a formal history ensemble. In a unitary description, two different histories may lead to the same reduced macroscopic state $(M,Q,S)$ while remaining different in hidden or environmental degrees of freedom. Adding a mass or energy label reduces the endpoint degeneracy, but it does not remove it in the ensembles studied here. The corrected interpretation separates dynamical entropy, state entropy, history entropy, and hidden conditional entropy. It also clarifies that the multinomial formula is a restricted history counting result rather than a proof of unique or identical microscopic \BH{} states.
\end{abstract}


\maketitle

\section{Introduction}

The thermodynamics of black holes (BHs) remains one of the central problems at the intersection of gravitation, statistical physics, and quantum theory. The Bekenstein proposal that \BHs{} carry entropy proportional to horizon area \cite{Bekenstein1973}, the mechanical laws of \BHs{} established by Bardeen, Carter, and Hawking \cite{BardeenCarterHawking1973}, and Hawking's derivation of \BH{} radiation \cite{Hawking1975} established the modern problem of explaining why a geometric object admits a thermodynamic description and how microscopic information is encoded in the entropy law. This question has been approached from many directions, including entanglement entropy \cite{Bombelli1986,Srednicki1993}, brick wall and near horizon state counting \cite{tHooft1985}, induced gravity and effective action methods \cite{FrolovNovikov1993,Solodukhin2011}, string theory microstate counting \cite{StromingerVafa1996}, loop gravity inspired counting \cite{Rovelli1996,KaulMajumdar2000}, and symmetry-based approaches to logarithmic corrections \cite{Carlip2000}. These approaches differ in microscopic interpretation, but they repeatedly point to the robustness of the leading area term and to the importance of subleading logarithmic corrections.

We use the following convention for a logarithmically corrected entropy:
\begin{equation}\label{eq:intro_log_entropy}
\frac{S}{\kb}=\frac{A}{4\lp^2}+\gamma\ln\!\left(\frac{A}{\lp^2}\right)+\text{const.}+\cdots,
\end{equation}
where $A$ is the horizon area and $\gamma$ is a dimensionless coefficient. In the discrete model studied below, the uncharged Schwarzschild result has the form
\begin{align}
\label{eq:intro_discrete_log_entropy}
\frac{S_{\mathrm{disc}}}{\kb}=N(A)
&=\frac{A}{16\pi K_{\mathrm{eff}}}
-\frac{1}{4}\ln\!\left(\frac{A}{A_0}\right) \nonumber \\
&\quad +\text{const.}+\cdots .
\end{align}
Here $A_0$ only makes the logarithm dimensionless. After matching the leading term to the Bekenstein--Hawking law, Eq.~\eqref{eq:intro_discrete_log_entropy} becomes a concrete version of Eq.~\eqref{eq:intro_log_entropy} with $\gamma=-1/4$ in the uncharged sector.

The one bit absorption picture proposed in Ref.~\cite{Blaschke2018} is a simple effective model, used to isolate a limited structural question. A \BH{} is built through a sequence of minimal absorption events, and each event is assigned one elementary unit of information. In the continuum limit, the resulting growth law reproduces the leading area dependence of the entropy. The advantage of the framework is that it is classical, transparent, and simple enough to allow exact control over the discrete recursion.

Several entropies can appear at once: the step count $N$, the thermodynamic state entropy $S_{\BH}(M,Q)$, the entropy of a set of possible histories, and the entropy produced by ignoring hidden data. If the exact microscopic state and the exact history were known, the corresponding uncertainty would vanish. If only the macroscopic parameters are kept, many histories or hidden states may become indistinguishable. The present paper therefore separates four quantities:
\begin{align}
S_{\mathrm{disc}}&=\kb N,\\
S_{\BH}&=\frac{\kb A}{4\lp^2},\\
S_{\mathrm{hist}}&=\kb\ln\Omega_{\mathrm{hist}},\\
S_{\mathrm{hid}|\mathrm{macro}}&=\kb\mathcal{H}(\mathfrak h|X),
\end{align}
where $\mathfrak h$ denotes the history label, $X$ denotes the retained macroscopic endpoint, and $\mathcal H$ denotes Shannon entropy. These objects may coincide only after a microscopic ensemble and its constraints have been specified.

We derive the corrected asymptotic expansion of the discrete growth law. For Schwarzschild--Tangherlini \BHs{} in $d$ spatial dimensions, the leading area-type term is straightforward, but the logarithmic coefficient must be extracted from the discrete recursion. The corrected coefficient is derived as
\begin{equation}\label{eq:intro_Bd_definition}
B_d=-\frac{1}{2(d-2)}.
\end{equation}
If this coefficient is assigned incorrectly, the interpretation of the asymptotic expansion becomes unstable.

The original one bit analysis used arbitrary dimension to test the universality of the uncharged logarithmic correction \cite{Blaschke2018}. We keep the Schwarzschild--Tangherlini calculation in general $d$ for that reason. The charged Reissner--Nordstr\"om calculation is performed in the physical four-dimensional RN geometry, corresponding to $d=3$ spatial dimensions in our notation.

We extend the model to the fixed charge RN sector. Charged \BHs{} test whether the logarithmic correction follows from the recursion or has been imported from the Schwarzschild case. We find that the fixed charge RN logarithmic coefficient is charge dependent. The mass step expansion already contains this dependence, and the asymptotic law for the step count inherits it.

The fixed charge calculation is not meant to describe the whole process by which a charged \BH{} is assembled. A realistic path may first create a charged seed with the final charge and then grow it by neutral absorptions. In the weakly charged macroscopic regime, the seed fixes only the lower cutoff of the later recursion, while the dominant entropy is generated during the neutral fixed charge stage.

The next objective is to correct the interpretation of charged history counting. If positive, negative, and neutral labels are all allowed, one can count ordered histories at fixed total step number and fixed final charge. This gives a Gaussian charge distribution and a leading history entropy $N\ln 3$. That result is the entropy of an unrestricted three-symbol alphabet.

We do not claim that minimal absorption history leading to a given endpoint is unique. The stronger statement compatible with quantum information conservation is more precise \cite{Preskill1992,Mathur2009}. Different absorption histories can lead to the same reduced macroscopic state when only $(M,Q,S)$ are observed, in the usual density-matrix sense of tracing over inaccessible degrees of freedom \cite{NielsenChuang2010}. They do not become the same exact quantum state. A unitary description should instead keep the missing data in hidden, environmental, or radiation degrees of freedom. In schematic form,
\begin{equation}\label{eq:intro_unitary_map}
|h_i\rangle\longrightarrow |M,Q,S\rangle_{\mathrm{macro}}\,|\eta_i\rangle_{\mathrm{hid}} .
\end{equation}
Thus, apparent degeneracy at the macroscopic level is a statement about coarse graining, not a statement that exact quantum states have been identified.

A deeper interpretation would require the local labels to be dynamically allowed stable absorption sectors rather than arbitrary symbols. In that stronger setting one would expect schematically
\begin{equation}\label{eq:stable_sector_entropy_intro}
S_{\mathrm{hist}}\sim\kb N\ln N_{\mathrm{stable}},
\end{equation}
where $N_{\mathrm{stable}}$ is the number of stable elementary absorption channels admitted by the effective theory. The present paper does not derive such a spectrum. This possibility is left as future work.

Once the discrete count and the history entropy are separated, the matching to the Bekenstein--Hawking law becomes clearer. The recursion reproduces the functional area law, but the overall normalization depends on the effective microscopic parameter in the growth step. Matching to the standard area law fixes that parameter. Therefore, the model should be viewed as an effective discrete growth picture with a microscopic scale that can be calibrated, not as a parameter free derivation of \BH{} thermodynamics.

The physical motivation for examining the charged sector also comes from a wider \BH{} context. Astrophysical \BHs{} are generally expected to remain nearly neutral because charge is efficiently discharged by the surrounding plasma and by pair creation processes \cite{Wald1974,Gibbons1975,Eardley1975,HiscockWeems1990}. If a simple discrete history model predicts that typical large \BHs{} are nearly neutral even before environmental discharge is added, then this is a useful consistency check. The combinatorial analysis below leads to this conclusion only within its formal ensemble.

The paper is organized as follows. Section~II reviews the discrete growth law and defines the quantities that are kept distinct. Section~III derives the corrected large mass asymptotics in general dimension. Section~IV studies the fixed charge RN sector and shows how the logarithmic coefficient changes with charge. Section~V reformulates the charged history combinatorics, adds the quantum coarse graining interpretation, and explains why Eq.~\eqref{eq:multinomial_histories} is a restricted history-counting formula. Section~VI discusses normalization. Section~VII gives numerical stability checks. The discussion and conclusion summarize the corrected interpretation.

\section{Discrete growth model}

We work in Planck units,
\[
G=c=\hbar=1.
\]
With this convention masses, lengths, and times are expressed in the same unit system. The fundamental dynamical variable is the number $N$ of absorption events. In the minimal one bit version of the model, one assigns
\begin{equation}\label{eq:Sdisc_def}
S_{\mathrm{disc}}=\kb N.
\end{equation}
This quantity is best understood as a dynamical bookkeeping entropy. It counts the number of elementary growth steps after one has decided that a single step carries one elementary absorption unit. Throughout the paper the elementary entropy unit is measured in natural logarithmic units, so that the phrase ``one bit'' denotes one elementary absorption unit with $S/\kb=N$. Restoring literal binary bits would replace $N$ by $N\ln 2$ and can be absorbed into the effective microscopic constant. The usefulness of writing the model in this way is that the large mass behavior of $N(M)$ can be studied directly and compared with the \BH{} area law. However, Eq.~\eqref{eq:Sdisc_def} is a postulate and not a consequence of combinatorics.

For a Schwarzschild--Tangherlini \BH{} in $d$ spatial dimensions, the discrete growth rule may be written as
\begin{align}
M_{N+1}&=M_N+\Delta M(M_N), \\
\Delta M(M)&=K_d\,M^{-\frac{1}{d-2}},
\end{align}
where $K_d>0$ is an effective parameter that contains the microscopic information about the wavelength condition imposed on the absorbed carrier. The first discrete value in the sequence defines the low-mass cutoff,
\begin{equation}
M_1\equiv M(N=1).
\end{equation}
All asymptotic formulas below are large mass expansions around this discrete recursion.

If one temporarily ignores the discreteness and replaces the difference equation by a differential equation, then
\begin{equation}
\frac{\dd M}{\dd N}=K_d\,M^{-\frac{1}{d-2}},
\end{equation}
which integrates to
\begin{equation}\label{eq:N0_general}
N_0(M)=\frac{M^p-M_1^p}{pK_d},
\qquad
p\equiv\frac{d-1}{d-2}.
\end{equation}
This leading term already has the correct area-type dependence, and this is the part emphasized in Ref.~\cite{Blaschke2018}. The main point of the present work is that the discrete problem contains more information than Eq.~\eqref{eq:N0_general}.

The model contains two separate calculational layers. The first is the dynamical recursion for $N(M)$ or $M(N)$. The second is the counting of different ordered histories that realize a given macroscopic state. The first layer is deterministic once the microscopic step rule is chosen. The second layer is probabilistic and combinatorial. This distinction is kept explicit at every stage.

\section{Corrected large mass asymptotics}

The discrete recursion can be inverted asymptotically with an Euler--Maclaurin analysis. For clarity and reproducibility, the short derivation is given explicitly in Appendix~\ref{app:general_d}. The resulting large mass expansion is
\begin{align}\label{eq:N_general_corrected}
N(M)=&\frac{M^p-M_1^p}{pK_d}-\frac{1}{2(d-2)}\ln\!\left(\frac{M}{M_1}\right) \nonumber
\\
&+c_0+\Order\!\left(M^{-\alpha}\right), 
\end{align}
with
\begin{equation}
\alpha>0.
\end{equation}
The detailed power of the first inverse-mass term depends on the sector, but the logarithmic coefficient does not. The coefficient is fixed by the discrete growth rule and equals
\begin{equation}\label{eq:Bd}
B_d=-\frac{1}{2(d-2)}.
\end{equation}

In equation~\eqref{eq:Bd} the logarithmic term does not arise from an arbitrary fitting ansatz, but comes from the mismatch between the exact recursion and its continuum approximation. In that sense, the logarithm is a genuine discreteness correction. Once one adopts the growth rule, the logarithmic coefficient is no longer adjustable.

For $d=3$, Eq.~\eqref{eq:N_general_corrected} reduces to Schwarzschild form
\begin{equation}\label{eq:N_schw_asym}
N(M)=\frac{M^2-M_1^2}{2K_3}-\frac{1}{2}\ln\!\left(\frac{M}{M_1}\right)+c_0+\Order(M^{-2}).
\end{equation}

\begin{figure}[t]
\centering
\includegraphics[width=0.95\linewidth]{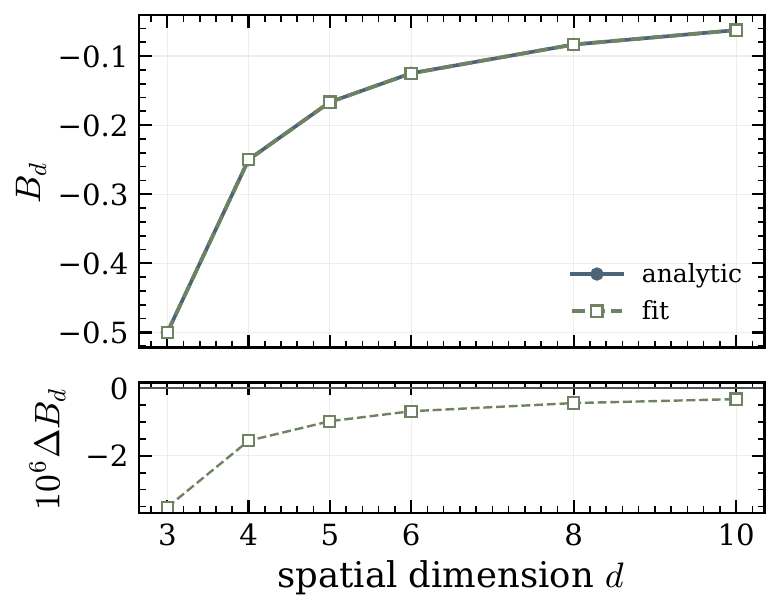}
\caption{Numerical verification of the corrected logarithmic coefficient in several dimensions. The extracted values follow the analytic prediction $B_d$. Where the analytic and fitted curves overlap, the residual inset shows the small difference; the residual values are rescaled for readability and the horizontal line marks zero.}
\label{fig:general_d_coefficients}
\end{figure}

The coefficient extraction across several dimensions is summarized in Table~\ref{tab:general_d}. The same trend is shown graphically in Fig.~\ref{fig:general_d_coefficients}. The fitted leading coefficient follows the continuum prediction, and the fitted logarithmic coefficient follows the corrected analytic value in Eq.~\eqref{eq:Bd}.

\begin{table}[t]
\centering
\begin{tabular}{c c c c c}
\toprule
$d$ & $A_d^{\rm ana}$ & $A_d^{\rm fit}$ & $B_d^{\rm ana}$ & $B_d^{\rm fit}$ \\
\midrule
3  & 0.25 & 0.25 & $-0.50$ & $-0.50$ \\
4  & 0.33 & 0.33 & $-0.25$ & $-0.25$ \\
5  & 0.38 & 0.38 & $-0.17$ & $-0.17$ \\
6  & 0.40 & 0.40 & $-0.13$ & $-0.13$ \\
8  & 0.43 & 0.43 & $-0.08$ & $-0.08$ \\
10 & 0.44 & 0.44 & $-0.06$ & $-0.06$ \\
\bottomrule
\end{tabular}
\caption{Analytic and fitted coefficients in the large mass expansion of the discrete recursion. The displayed values are rounded to two decimals.}
\label{tab:general_d}
\end{table}

Once Eq.~\eqref{eq:N_general_corrected} is established, the later charged analysis can be treated as a controlled deformation of the recursion rather than as an unrelated numerical calculation.

\section{Fixed charge Reissner--Nordstr\"om}

We now turn to the Reissner--Nordstr\"om \BH{} at fixed charge $Q$. Here fixed charge means that $Q$ is held as a macroscopic parameter during the mass recursion: the absorption step changes $M$, while no charge fluctuation or discharge term is included. This is different from the formal charged-history ensemble discussed in Sec.~V.

The outer horizon radius is
\begin{equation}
r_+(M,Q)=M+\sqrt{M^2-Q^2}.
\end{equation}
For a massless carrier whose wavelength is proportional to the horizon size, the discrete mass increment takes the form
\begin{equation}\label{eq:deltaM_RN_exact}
\Delta M(M,Q)=\frac{K_{\mathrm{eff}}}{r_+(M,Q)}.
\end{equation}
This expression is exact within the assumptions of the model.

The fixed charge RN recursion should be distinguished from the charged
history ensemble: in the charge calculation $Q$ is held fixed, so the
recursion probes how a charged background modifies the mass step. A literal
interpretation of the charged history labels $\{+,-,0\}$, however, would
require idealized charged elementary absorption channels. Since the Standard
Model contains no known charged massless bosons, such labels should be viewed
as formal effective channels rather than as realistic particle physics
degrees of freedom \cite{WeinbergQFT2,PeskinSchroeder}. We also assume that the carrier energy is fixed only by the wavelength relation and neglect possible electromagnetic interaction terms associated with the horizon electrostatic potential $\Phi_H=Q/r_+$. In a full RN absorption problem, such terms could affect both the minimum absorbed energy and the set of allowed charged trajectories \cite{Wald1984,ChristodoulouRuffini1971,BardeenCarterHawking1973}.

A useful way to interpret the fixed charge recursion is to separate the construction of a final state $(M_f,Q_f)$ into two stages. The first stage creates a charged seed with charge $Q_f$ and mass
\begin{equation}\label{eq:seed_mass_scaling}
    M_{\rm seed}(Q_f)\sim {\cal O}(|Q_f|).
\end{equation}
The extremality condition gives the lower bound $M_{\rm seed}\ge |Q_f|$, while electrostatic work and details of charged absorption may change the numerical coefficient. This charged stage is not modeled by Eq.~\eqref{eq:deltaM_RN_exact}. It fixes the initial data for the later recursion.

The second stage keeps the charge fixed and grows the seed by neutral minimal absorptions,
\begin{equation}\label{eq:neutral_growth_after_seed}
    M_{n+1}=M_n+\frac{K_{\mathrm{eff}}}{r_+(M_n,Q_f)} .
\end{equation}
For a macroscopic weakly charged final state, $M_f\gg |Q_f|$, the entropy associated with the seed is of order
\begin{equation}\label{eq:seed_entropy_scaling}
    S_{\rm seed}\sim Q_f^2,
\end{equation}
whereas the final entropy scales as
\begin{equation}
    S_f\sim M_f^2 .
\end{equation}
Therefore
\begin{equation}\label{eq:seed_entropy_ratio}
    \frac{S_{\rm seed}}{S_f}
    \sim
    \left(\frac{Q_f}{M_f}\right)^2 .
\end{equation}
Thus, for $M_f\gg |Q_f|$, almost all of the entropy is produced during the neutral fixed charge growth stage. The complicated physics of charged absorption changes the seed and the low mass cutoff, but it does not change the large mass asymptotic structure of the neutral recursion. This argument does not apply to near extremal endpoints, where $M_f$ is comparable to $|Q_f|$ and the seed contribution is not parametrically small.

The large mass expansion of the step is
\begin{align}
\label{eq:deltaM_RN_series}
\Delta M(M,Q)=&K_{\mathrm{eff}}\bigg[
\frac{1}{2M}
+\frac{Q^2}{8M^3} \nonumber \\
&+\frac{Q^4}{16M^5}
+\Order(M^{-7})
\bigg].
\end{align}
The first two coefficients are numerically well resolved.

It is convenient to rewrite Eq.~\eqref{eq:deltaM_RN_series} as
\begin{equation}
\Delta M(M,Q)=\frac{c_1}{M}+\frac{c_3}{M^3}+\frac{c_5}{M^5}+\Order(M^{-7}),
\end{equation}
with
\begin{equation}
c_1=\frac{K_{\mathrm{eff}}}{2},
\qquad
c_3=\frac{K_{\mathrm{eff}}Q^2}{8},
\qquad
c_5=\frac{K_{\mathrm{eff}}Q^4}{16}.
\end{equation}
Thus $c_1$, $c_3$, and $c_5$ are the coefficients of the successive inverse mass terms. For the numerical choice $K_{\mathrm{eff}}=10$ and $Q=30$, a direct fit of the exact expression \eqref{eq:deltaM_RN_exact} over a large mass window gives
\begin{align}
c_1^{\rm fit} &= 5.00, & c_1^{\rm ana} &= 5.00, \\
c_3^{\rm fit} &= 1124.65, & c_3^{\rm ana} &= 1125.00.
\end{align}
The coefficient $c_5$ is more sensitive to the fitting window and is therefore not used as a primary quantitative test. This does not indicate any problem with the asymptotic expansion itself. The higher order coefficients are harder to extract reliably from a finite mass range.

\begin{figure*}[!t]
\centering
\includegraphics[width=0.96\textwidth]{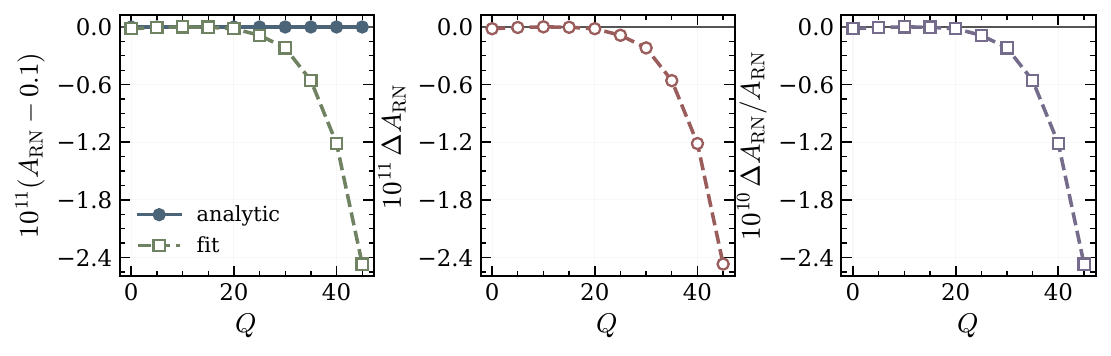}
\caption{RN leading coefficient \(A\) and its numerical deviations as functions
of charge. Panel (a) shows the fitted and analytic values of \(A_{\rm RN}\);
the analytic prediction is the charge-independent value \(A_{\rm RN}=1/10\).
Panel (b) shows the absolute deviation
\(\Delta A=A_{\rm fit}-A_{\rm analytic}\). Panel (c) shows the relative
deviation
\((A_{\rm fit}-A_{\rm analytic})/A_{\rm analytic}\). The deviations remain
small over the tested charge range and are consistent with numerical fitting
error.}
\label{fig:rn-A-combined}
\end{figure*}

The fixed charge asymptotic relation for the step count is obtained by integrating the inverse step and then adding the discrete correction. The corresponding calculation is shown in Appendix~\ref{app:rn_derivation}. The result is
\begin{align}\label{eq:N_RN_corrected}
N(M,Q)=&\frac{M^2-M_1^2}{K_{\mathrm{eff}}}
-\left(\frac{1}{2}+\frac{Q^2}{2K_{\mathrm{eff}}}\right)
\ln\!\left(\frac{M}{M_1}\right) \nonumber \\
&+C_0(Q)+C_2(Q)M^{-2}+\Order(M^{-4}).
\end{align}
We separate the coefficient of the quadratic term and write
\begin{equation}
A_{\mathrm{RN}}=\frac{1}{K_{\mathrm{eff}}}.
\end{equation}
For the illustrative numerical choice $K_{\mathrm{eff}}=10$, a direct fit gives
\begin{equation}
A_{\mathrm{RN}}^{\rm fit}=0.10,
\end{equation}
in agreement with Eq.~\eqref{eq:N_RN_corrected}. The charge independence of the leading quadratic coefficient is checked in Fig.~\ref{fig:rn-A-combined}, where the fitted values and their numerical deviations are shown.

\begin{figure}[t]
\centering
\includegraphics[width=0.95\linewidth]{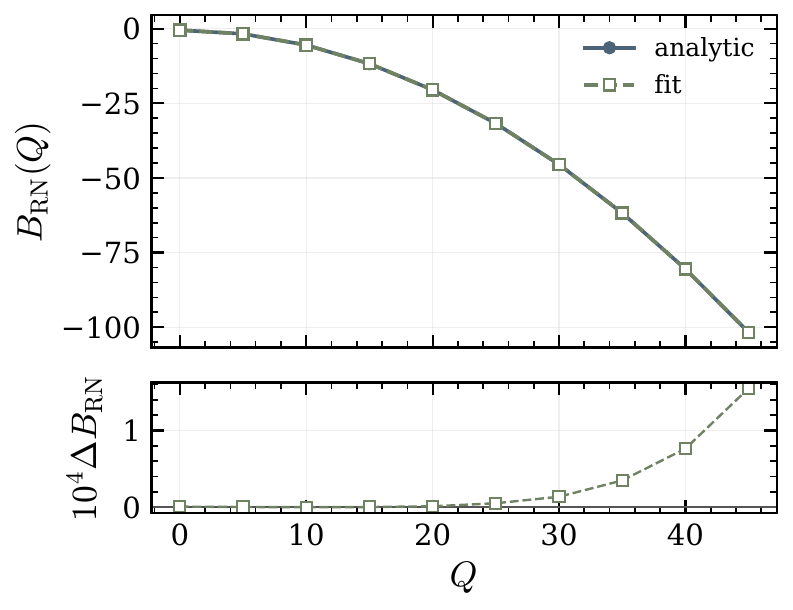}
\caption{Charge dependence of the RN logarithmic coefficient. The numerical data follow the analytic law $B_{\mathrm{RN}}(Q)=-1/2-Q^2/(2K_{\mathrm{eff}})$.}
\label{fig:blog_vs_charge}
\end{figure}

\begin{figure}[t]
\centering
\includegraphics[width=0.95\linewidth]{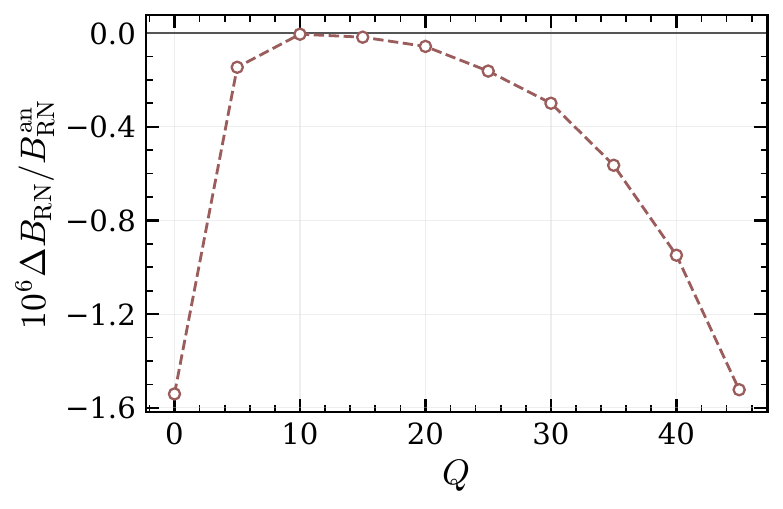}
\caption{Relative deviation between the fitted RN logarithmic coefficient and the analytic prediction. The visible discrepancies are small and decrease in the asymptotic regime.}
\label{fig:blog_rel_dev}
\end{figure}

So the logarithmic coefficient in the RN sector is different from the Schwarzschild value. Instead,
\begin{equation}\label{eq:RN_log_coeff}
B_{\mathrm{RN}}(Q)=-\frac{1}{2}-\frac{Q^2}{2K_{\mathrm{eff}}}.
\end{equation}
The charge dependence is therefore explicit and unavoidable. For example, this gives $-1.75$ at $Q=5.00$, $-5.50$ at $Q=10.00$, and $-20.50$ at $Q=20.00$. The full charge dependence of the fitted logarithmic coefficient is shown in Fig.~\ref{fig:blog_vs_charge}, while the relative deviation from the analytic prediction is displayed in Fig.~\ref{fig:blog_rel_dev}.

The RN values show that the model is sensitive enough to distinguish asymptotic sectors and to assign them different logarithmic structures, instead of being a small perturbation of the Schwarzschild variant.

\section{Combinatorics of charged histories}

We now separate the deterministic recursion from a distinct combinatorial question. This section does not count the unique sequence generated by a chosen minimal protocol. It counts ordered charge sign histories in a formal ensemble, where each entry records only whether the absorbed channel is positive, negative, or neutral. It does not specify particle species, energy, internal quantum state, emission correlations, radiation, or environmental degrees of freedom.

Consider an ordered sequence
\begin{equation}
H=(h_1,\dots,h_N),
\qquad
h_i\in\{+, -, 0\}.
\end{equation}
If the occupation numbers are $(N_+,N_-,N_0)$, the number of ordered sign histories is
\begin{equation}\label{eq:multinomial_histories}
\Omega(N_+,N_-,N_0)=\frac{N!}{N_+!\,N_-!\,N_0!},
\end{equation}
with
\begin{equation}
N=N_++N_-+N_0,
\qquad
Q=\ech(N_+-N_-).
\end{equation}
The associated history entropy is
\begin{equation}
S_{\mathrm{hist}}=\kb\ln\Omega .
\end{equation}
This is a history entropy for a restricted alphabet. It is not yet the thermodynamic state entropy of a \BH{}.

Equation~\eqref{eq:multinomial_histories} describes the larger ensemble in which many ordered sign strings are allowed and only a few final labels are retained. Therefore the formula should not be used as a proof that the minimal absorption history is unique.

The reason is the distinction between exact quantum states and reduced macroscopic states. Suppose two histories $h_i$ and $h_j$ lead to the same observed endpoint $(M,Q,S)$. A unitary description should not identify the corresponding exact states. Instead one should write schematically
\begin{align}\label{eq:unitary_hidden_map}
|h_i\rangle &\longrightarrow |M,Q,S\rangle_{\mathrm{macro}}|\eta_i\rangle_{\mathrm{hid}}, \nonumber
\\
|h_j\rangle &\longrightarrow |M,Q,S\rangle_{\mathrm{macro}}|\eta_j\rangle_{\mathrm{hid}},
\end{align}
with $\langle\eta_i|\eta_j\rangle=0$ for perfectly distinguishable hidden records. Then
\begin{align}\label{eq:trace_distances}
D_{\mathrm{full}}&=\frac{1}{2}\left\lVert\rho^{(i)}_{\mathrm{full}}-\rho^{(j)}_{\mathrm{full}}\right\rVert_1=1, \nonumber
\\
D_{\mathrm{macro}}&=\frac{1}{2}\left\lVert\rho^{(i)}_{\mathrm{macro}}-\rho^{(j)}_{\mathrm{macro}}\right\rVert_1=0,
\end{align}
where $\rho_{\mathrm{macro}}=\operatorname{Tr}_{\mathrm{hid}}\rho_{\mathrm{full}}$. Thus the two histories become identical only for an observer who has traced over the hidden labels. This is the standard reduced-state construction used in quantum information theory \cite{NielsenChuang2010}. It is compatible with quantum information conservation. The apparent degeneracy is a coarse grained degeneracy. It can be illustrated by a state function model, let each event add one entropy unit $s_0$ and one charge label $q_i\in\{-1,0,1\}$. Then
\begin{align}\label{eq:toy_state_function}
S_f&=S_0+Ns_0, 
\\
Q_f&=e_0\sum_{i=1}^{N}q_i, 
\\
M_f&=M_{\mathrm{RN}}(S_f,Q_f),
\end{align}
where, in four-dimensional RN units,
\begin{equation}\label{eq:rn_mass_from_entropy}
M_{\mathrm{RN}}(S,Q)=\frac{1}{2}\left(r_++\frac{Q^2}{r_+}\right),
\qquad
r_+=\sqrt{\frac{S}{\pi}} .
\end{equation}
All histories with the same $\sum_iq_i$ have the same final $(S_f,Q_f,M_f)$ in this toy model. Therefore minimal history uniqueness is not a generic consequence of the state variables alone. It requires an additional canonical protocol.

\begin{figure}[t]
\centering
\includegraphics[width=0.95\linewidth]{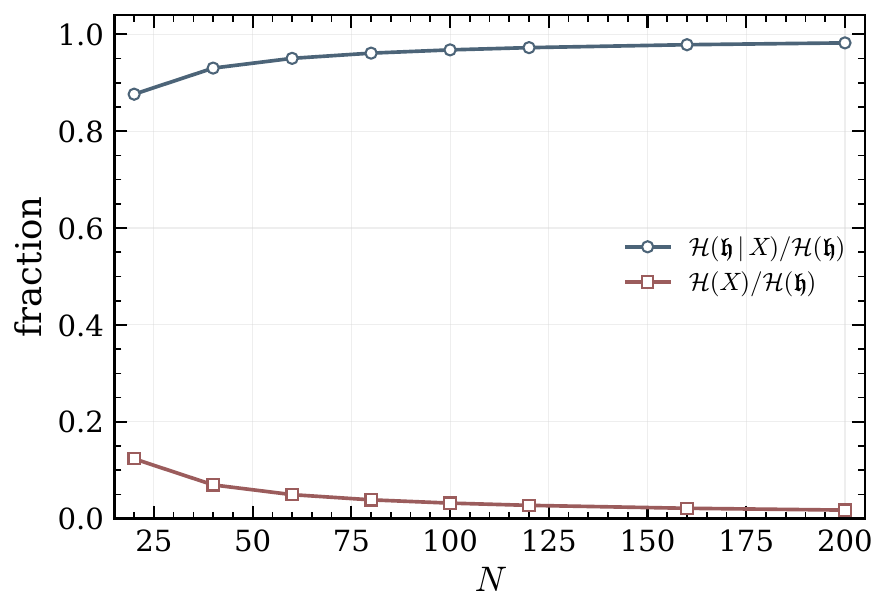}
\caption{Fraction of total history entropy stored in the macroscopic charge endpoint $X$ and in the hidden conditional sector $\mathcal{H}(\mathfrak h|X)$. The unrestricted sign history entropy grows as $N\ln3$, while the macro charge entropy grows much more slowly. Most history information is therefore hidden when only the endpoint is retained.}
\label{fig:hidden_information_fraction}
\end{figure}

The corresponding entropy decomposition is
\begin{equation}\label{eq:entropy_decomposition}
\mathcal{H}(\mathfrak h)=\mathcal{H}(X)+\mathcal{H}(\mathfrak h|X),
\end{equation}
where $\mathfrak h$ is the full history label, $X$ is the retained macroscopic endpoint, and $\mathcal H$ is Shannon entropy \cite{Shannon1948}. For the unrestricted sign alphabet, $\mathcal{H}(\mathfrak h)=N\ln3$. The macro entropy $\mathcal{H}(X)$ grows only logarithmically with the number of charge sectors, while $\mathcal{H}(\mathfrak h|X)$ contains the information hidden by the coarse graining. Figure~\ref{fig:hidden_information_fraction} shows this behavior for the exact dynamic programming count.

At fixed $N$ and fixed $Q$, the saddle-point evaluation gives the result derived in Appendix~\ref{app:history_derivation}:
\begin{align}\label{eq:Shist_main}
S_{\mathrm{hist}}(N,Q)=\kb\bigg[N\ln 3-\frac{3}{4}\frac{Q^2}{\ech^2N} \nonumber \\-\frac{1}{2}\ln N+\Order(1)\bigg].
\end{align}
The leading slope is $\ln 3\approx1.10$ per step. Unrestricted counting of positive, negative, and neutral sign histories does not reproduce one elementary unit per step. It gives the Shannon entropy of a ternary alphabet. Hence the one bit rule is an additional microscopic normalization assumption unless the ensemble is restricted by additional dynamics.

\begin{figure*}[t]
\centering
\includegraphics[width=\textwidth]{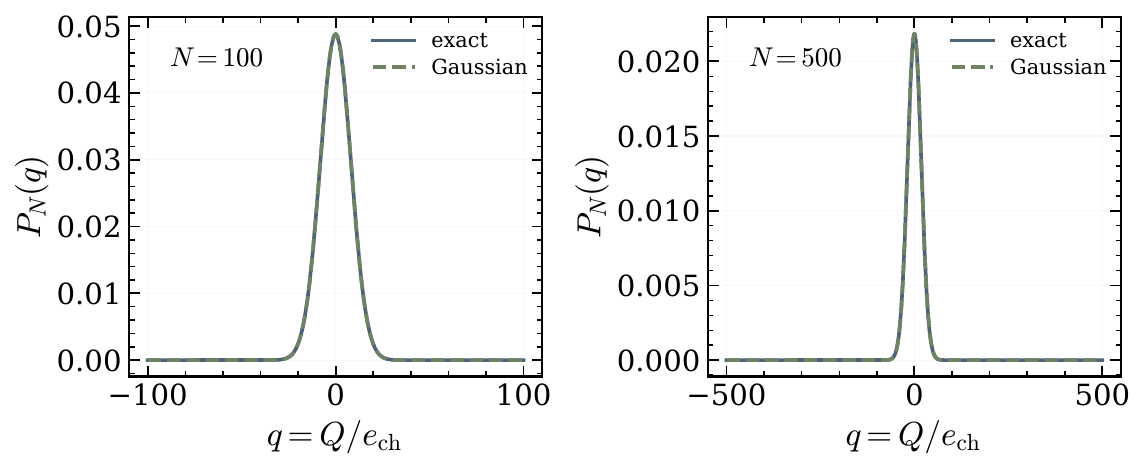}
\caption{Charge distribution from formal sign-history counting at $N=100$ and $N=500$, compared with the Gaussian approximation in Eq.~\eqref{eq:charge_gaussian}. The agreement improves with increasing $N$.}
\label{fig:charge_distribution}
\end{figure*}

The fixed-$N$ charge distribution follows from Eq.~\eqref{eq:Shist_main}:
\begin{equation}\label{eq:charge_gaussian}
P(Q\mid N)\propto\exp\!\left[-\frac{3}{4}\frac{Q^2}{\ech^2N}\right].
\end{equation}
This is a Gaussian with variance
\begin{equation}\label{eq:variance}
\sigma_Q^2=\frac{2}{3}\ech^2N .
\end{equation}
Hence the typical charge grows only as $\sqrt{N}$, while the total number of steps grows as $N$. It follows that
\begin{equation}
\frac{|Q|}{\ech N}\sim N^{-1/2},
\end{equation}
so large \BHs{} in this ensemble are typically close to neutrality.
The exact fixed-$N$ charge distributions and their Gaussian approximations are shown in Fig.~\ref{fig:charge_distribution}.

\begin{figure}[t]
\centering
\includegraphics[width=0.95\linewidth]{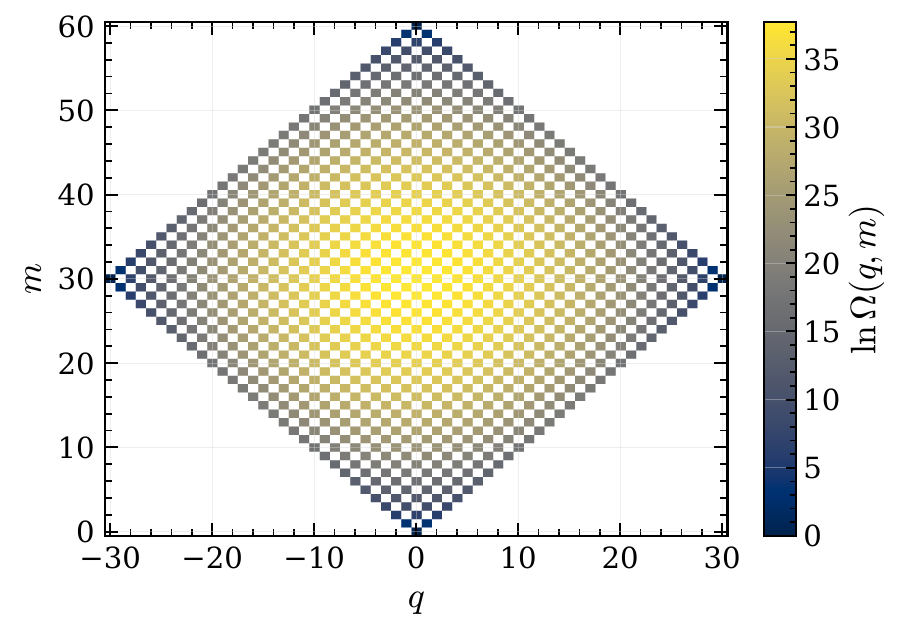}
\caption{Endpoint degeneracy in a toy ensemble with both charge and mass labels. Adding the mass label separates more endpoints than charge alone, but it does not make the endpoint map one to one. The plotted quantity is $\ln\Omega(q_{\mathrm{sum}},m_{\mathrm{sum}})$ for $N=30$.}
\label{fig:mass_charge_endpoint_degeneracy}
\end{figure}

Adding a mass or energy label does not remove the degeneracy. For example, consider four channels with labels
\begin{equation}\label{eq:mass_charge_channels}
(q,m)\in\{(0,0),(1,1),(-1,1),(0,2)\} .
\end{equation}
The final endpoint is now $(q_{\mathrm{sum}},m_{\mathrm{sum}})$ instead of charge alone. The number of endpoints grows, and the degeneracy is reduced. But many ordered histories still share the same endpoint. Figure~\ref{fig:mass_charge_endpoint_degeneracy} shows the exact count for $N=30$.

If the local mass increment depends on the instantaneous value of $M$ and $Q$, then the order of the same labels can also change the final mass. In that case history entropy and state entropy are even less automatic to identify. With finite macroscopic resolution, however, many histories can still fall in the same measured endpoint bin. The safe conclusion is therefore limited -- Eq.~\eqref{eq:multinomial_histories} is a restricted history counting formula.

In realistic astrophysical settings, \BHs{} are expected to discharge efficiently through their environment and through pair creation processes \cite{Wald1974,Gibbons1975,Eardley1975,HiscockWeems1990}. The present calculation is not a model of that discharge. It only shows that the unrestricted sign history ensemble already favors small charge-to-step ratios at large $N$.

The history-counting result therefore has a restricted role. It separates deterministic growth dynamics, formal history multiplicity, and macroscopic coarse graining. It also shows why the phrase ``degeneracy'' must be used carefully. In this section it means many histories per reduced macroscopic endpoint, not identity of exact quantum states.

\section{Logarithmic term and normalization}

In the Schwarzschild sector, Eq.~\eqref{eq:N_schw_asym} implies the leading large mass behavior
\begin{equation}
N(M)\simeq\frac{M^2}{K_{\mathrm{eff}}},
\end{equation}
where we rename the four-dimensional constant as $K_{\mathrm{eff}}$ for direct comparison with the RN formulas. In this convention the two constants are related by $K_{\mathrm{eff}}=2K_3$, because the RN step reduces to $\Delta M=K_{\mathrm{eff}}/(2M)$ when $Q=0$. Since the horizon area is
\begin{equation}
A=16\pi M^2,
\end{equation}
we immediately obtain
\begin{equation}\label{eq:N_of_A}
N(A)=\frac{A}{16\pi K_{\mathrm{eff}}}-\frac{1}{4}\ln\!\left(\frac{A}{A_0}\right)+\text{const.}+\cdots.
\end{equation}
Thus the discrete model reproduces the familiar structural form of \BH{} entropy: a leading area term and a subleading logarithmic correction.

Equation~\eqref{eq:N_of_A} is important because it separates two issues that are often merged too early. The first issue is functional dependence. At that level the model succeeds as the recursion produces the correct area scaling and the expected kind of subleading correction. The second issue is normalization. At that level the model still contains the free effective parameter $K_{\mathrm{eff}}$. The recursion therefore does not predict the full Bekenstein--Hawking coefficient until one fixes the microscopic scale.

If one now identifies the dynamical entropy assignment with the thermodynamic entropy,
\begin{equation}
S_{\mathrm{disc}}=\kb N\simeq\frac{\kb}{16\pi K_{\mathrm{eff}}}A,
\end{equation}
then matching to
\begin{equation}
S_{\mathrm{BH}}=\frac{\kb A}{4\lp^2}
\end{equation}
requires
\begin{equation}\label{eq:Keff_match}
K_{\mathrm{eff}}=\frac{\lp^2}{4\pi}.
\end{equation}

\begin{figure}[t]
\centering
\includegraphics[width=0.95\linewidth]{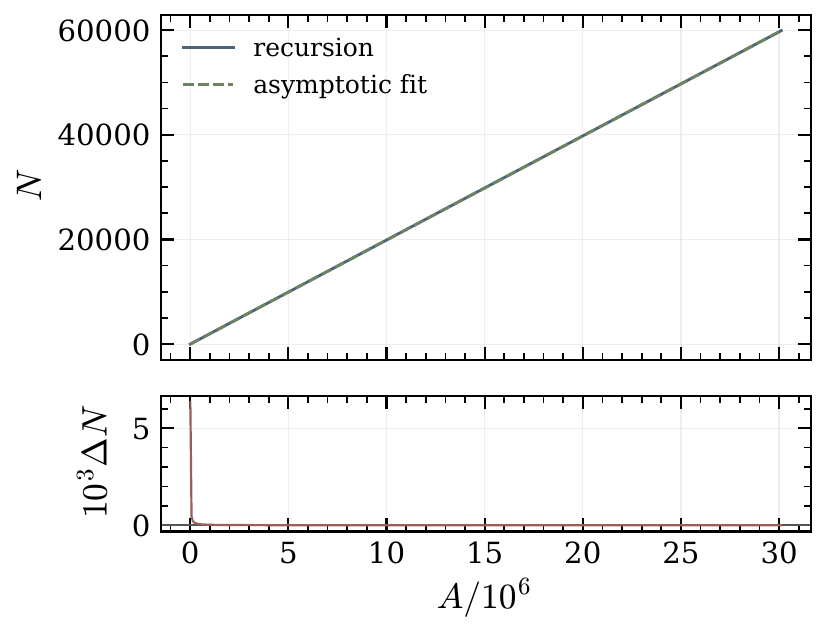}
\caption{Area-law scaling of the discrete step count in the Schwarzschild sector. The fit follows the form $N(A)=A/(16\pi K_{\mathrm{eff}})-(1/4)\ln(A/A_0)+\cdots$.}
\label{fig:area_scaling}
\end{figure}

For our setup with $K_{\mathrm{eff}}=10.00$, the fitted area coefficient is
\begin{equation}
\frac{1}{16\pi K_{\mathrm{eff}}}=1.99\times 10^{-3},
\end{equation}
in agreement with the analytic value. The logarithmic coefficient is likewise consistent with $-0.25$ when the asymptotic formula is written in terms of area rather than mass, as shown in Fig.~\ref{fig:area_scaling}.

\begin{figure}[t]
\centering
\includegraphics[width=0.95\linewidth]{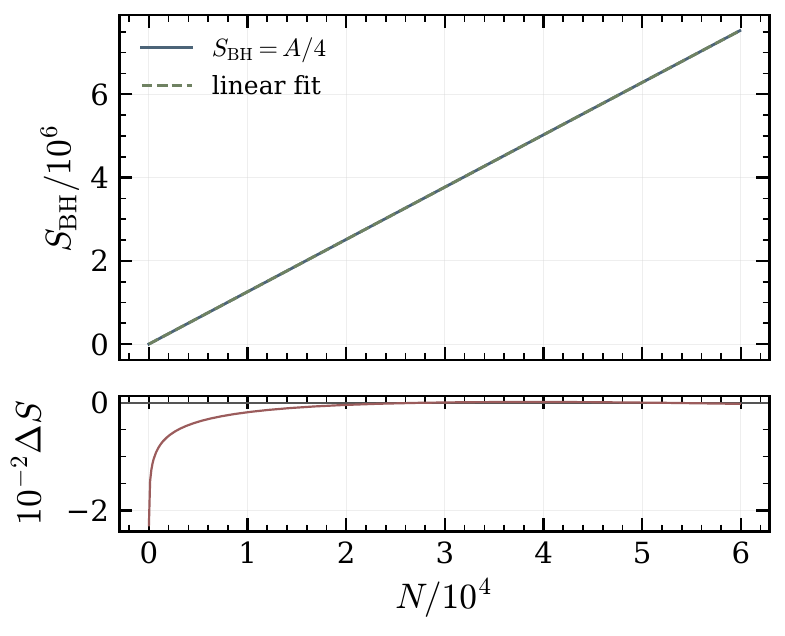}
\caption{Linear calibration between the discrete step count and the standard Schwarzschild entropy for the numerical parameter choice used in the notebook. The slope depends on the chosen microscopic scale.}
\label{fig:bh_calibration}
\end{figure}

For the same numerical setup, a direct calibration against the standard Schwarzschild entropy gives
\begin{equation}
S_{\mathrm{BH}}\approx125.66\,N+1484.07.
\end{equation}
This number is setup dependent. It reflects the chosen value of $K_{\mathrm{eff}}$ and the finite fitting interval. What matters conceptually is not the particular numerical slope but the fact that the recursion admits a clean linear calibration once the microscopic scale is fixed, as illustrated in Fig.~\ref{fig:bh_calibration}.

A side result of the numerical exploration is that a massive carrier version of the model falls outside the same asymptotic universality class. Once the carrier rest mass dominates the energy increment, the step tends to a constant rather than to $1/M$, and the area law is lost. For that reason, the present paper restricts the main analysis to the massless or ultrarelativistic sector.

\section{Numerical stability checks}

\begin{figure}[htbp]
\centering
\includegraphics[width=0.45\textwidth]{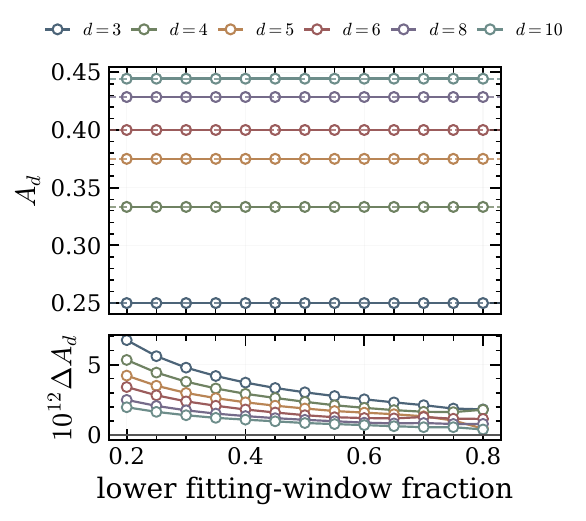}
\caption{Window stability of the leading coefficient \(A\). The fitted values
remain constant as the lower fitting-window fraction is varied.}
\label{fig:window-stability-A}
\end{figure}

\begin{figure}[htbp]
\centering
\includegraphics[width=0.45\textwidth]{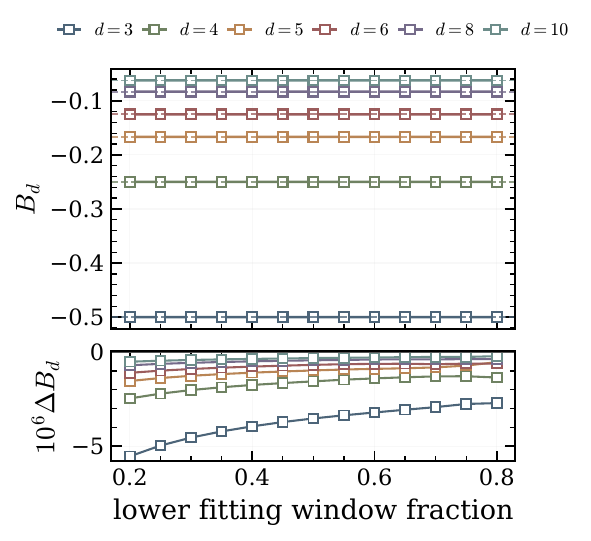}
\caption{Window stability of the logarithmic coefficient \(B\). The fitted
values remain stable under changes of the fitting window and reproduce the
analytic coefficient.}
\label{fig:window-stability-B}
\end{figure}

To test whether the extracted coefficients are robust, the fitting window was
varied. The fitted leading coefficient \(A\) is stable under changes of the
lower fitting-window fraction. For the dimensions shown, the extracted values
agree with
\begin{equation}
    A_d=\frac{d-2}{2(d-1)}.
\end{equation}
The fitted logarithmic coefficient \(B\) is likewise stable and agrees with
\begin{equation}
    B_d=-\frac{1}{2(d-2)}.
\end{equation}
This confirms that the fitted coefficients are not artifacts of a particular
choice of fitting window. Figures~\ref{fig:window-stability-A} and~\ref{fig:window-stability-B} show the corresponding stability checks.

For the RN charge dependent test with \(K_{\mathrm{eff}}=10\), the logarithmic
coefficient follows the analytic form
\begin{equation}
    B_{\rm RN}(Q)
    =
    -\frac{1}{2}
    -\frac{Q^2}{20}.
\end{equation}
The numerical fit reproduces this dependence over the tested charge range, as shown in Fig.~\ref{fig:blog_vs_charge}.
The charge contribution appears in the logarithmic coefficient rather than in the leading extensive normalization.

For the same numerical choice, the leading RN coefficient \(A\) is independent
of the charge and is given by \(A_{\rm RN}=1/10\). The direct plot of
\(A_{\rm fit}\) against \(Q\) shows only a very small numerical displacement
from the analytic horizontal line. We separately plot the
absolute and relative deviations
\begin{equation}
    \Delta A
    =
    A_{\rm fit}-A_{\rm analytic},
\end{equation}
and
\begin{equation}
    \frac{\Delta A}{A_{\rm analytic}}
    =
    \frac{A_{\rm fit}-A_{\rm analytic}}{A_{\rm analytic}}.
\end{equation}
The deviations are at the level of numerical fitting error and remain very small
over the full charge range tested.

\section{Discussion}\label{sec:discussion}

The results of this work place the discrete one bit growth model in a clearer and more restricted theoretical setting. A genuinely discrete recursion already contains nontrivial information about \BH{} thermodynamics. In particular, it fixes the leading large mass structure of the step count and generates both the area term and the first logarithmic correction without requiring a full microscopic state counting construction. This makes the framework conceptually close to other discrete approaches to \BH{} physics, especially those motivated by horizon area quantization and information bookkeeping \cite{Bekenstein1973,BekensteinMukhanov1995,Blaschke2018}.

It does not derive \BH{} microstates from an underlying quantum theory, and it should not be presented as an alternative to such derivations. In loop based, conformal, string inspired, and thermal fluctuation approaches, logarithmic corrections arise from microscopic degeneracy formulas, near horizon symmetry arguments, or statistical fluctuations around equilibrium \cite{KaulMajumdar2000,Carlip2000,DasMajumdarBhaduri2002,Sen2013}. Here the origin of the logarithm comes from the asymptotic inversion of a discrete growth law. Thus, the logarithmic term found here is best interpreted as a quasiclassical discreteness effect rather than as a full quantum gravitational state counting result. The model provides a clean benchmark for separating corrections that already follow from discrete kinematics from those that require genuine microscopic quantum input.

The fixed charge RN sector illustrates this point especially well. Once charge is held fixed during the large mass expansion, the logarithmic coefficient is no longer the Schwarzschild one. Instead it acquires an explicit $Q$ dependence. This shows that the logarithmic term is sector sensitive already at the level of the discrete classical recursion. Any attempt to extend discrete growth models to rotating, charged, or more general stationary \BHs{} must allow the logarithmic structure to depend on the macroscopic sector under consideration rather than assuming a single universal Schwarzschild type coefficient from the start. In that sense, the RN calculation is not merely an extension of the model, but a test of whether the discrete framework remains logically consistent once additional conserved quantities are introduced.

A second important outcome concerns the meaning of entropy in the model. The dynamical assignment $S_{\mathrm{disc}}=\kb N$ and the combinatorial entropy of ordered histories are not the same object. The multinomial counting of charged and neutral histories leads to a leading entropy proportional to $N\ln 3$, whereas the one bit rule gives $\kb N$ by assumption. This distinction should be stated explicitly because it determines the interpretive status of the framework. If one wants to keep the one bit rule, then it must be regarded as an additional microscopic postulate, not as a consequence of unrestricted counting. In the present paper the multinomial ensemble is a formal benchmark.

A more expansive interpretation would require the three labels to correspond to dynamically realizable stable absorption sectors. Then the leading history entropy would probe the number of stable local sectors, schematically $S_{\mathrm{hist}}\sim\kb N\ln N_{\mathrm{stable}}$, rather than the size of an arbitrary alphabet. This possibility is naturally related to Wheeler's information based ``it from bit'' viewpoint, Bekenstein and Mukhanov's discrete \BH{} spectrum, 't Hooft's dimensional reduction perspective, Carlip's near horizon conformal counting, Wheeler's geons, and the wider language of solitons, topological sectors, and semiclassical channels \cite{Wheeler1990,BekensteinMukhanov1995,tHooft1993,Carlip2000,Wheeler1955,MantonSutcliffe2004,Rajaraman1982}. These references motivate the possibility that a small number of stable channels could have physical meaning.

The RN extension also rests on idealized physical assumptions. First, it effectively uses charged massless carriers. Since the Standard Model contains no known charged massless bosons, this assumption is best viewed as a formal probe of the recursion rather than as a realistic microscopic matter model \cite{WeinbergQFT2,PeskinSchroeder}. Second, the mass step assumes that the carrier energy is governed by a wavelength scale alone. In a charged RN background, the electrostatic potential $\Phi_H=Q/r_+$ and Coulomb interaction energy can affect the absorption process \cite{Wald1984,ChristodoulouRuffini1971,BardeenCarterHawking1973}. Third, the paper assumes one fixed elementary information unit per absorption event. For charged carriers this remains a nontrivial assumption, because electromagnetic interactions may suppress or forbid many histories that are counted in the unrestricted multinomial ensemble.

The seed plus neutral growth argument limits the effect of this idealization. If the final state is weakly charged, $M_f\gg |Q_f|$, the charged construction contributes only an initial entropy of order $Q_f^2$, while the final entropy is of order $M_f^2$. The fixed charge RN recursion should therefore be interpreted as the asymptotic neutral growth regime above a charged seed, not as a complete microscopic model of charged accretion.

The area law itself is recovered only at the level of functional dependence until the microscopic parameter is calibrated. The normalization depends on $K_{\mathrm{eff}}$, and matching to the Bekenstein--Hawking law fixes that parameter. It places the model in the broader class of effective descriptions in which the large scale structure is robust but the overall normalization must be supplied by external physical input. From this perspective, the framework may be useful as a simplified laboratory for studying how discrete growth rules encode thermodynamic behavior before one commits to a more complete underlying theory. It may also be valuable in settings where one wants to compare the standard area law with generalized entropy formulas, for example in higher curvature theories where the relevant thermodynamic quantity is the Wald entropy rather than $A/4$ \cite{Wald1993,IyerWald1994}.

Several directions for further work follow naturally from the present results. The first is to extend the analysis to rotating and Kerr--Newman sectors, where multiple macroscopic parameters are present and the asymptotic structure may be more stable. The second is to replace the formal charged alphabet by an explicit dynamical model of allowed absorption channels. Such a model should also describe the seed creation stage and the near extremal regime, where the seed entropy is not negligible. The third is to examine whether modified elementary absorption rules can produce different universality classes for the logarithmic term. The fourth is to embed the discrete recursion into more general gravitational settings, such as higher curvature theories or effective models with nonstandard horizon thermodynamics, in order to test which features of the present construction survive beyond Einstein gravity. Finally, the distinction between dynamical entropy and history entropy suggests that the model may also serve as a useful toy framework for clarifying how theoretical assumptions enter \BH{} thermodynamics.

The overall conclusion is therefore moderate but positive. The discrete one bit model should not be interpreted as a complete microscopic derivation of \BH{} entropy. However, it does provide a transparent quasiclassical framework in which area scaling, logarithmic corrections, charge fluctuations, and normalization issues can all be studied in a controlled way. In that limited but well defined sense, it can serve as a useful bridge between classical horizon kinematics and more microscopic approaches to \BH{} thermodynamics.

\section{Conclusion}

We have reformulated the one bit \BH{} growth as a discrete problem and derived its corrected asymptotic structure. The Schwarzschild--Tangherlini recursion in $d$ spatial dimensions gives a logarithmic correction. The fixed charge RN sector gives the different coefficient and the leading quadratic coefficient $1/K_{\mathrm{eff}}$. The discrete recursion reproduces the functional area law after matching the dynamical entropy assignment to the Bekenstein--Hawking normalization. For weakly charged macroscopic endpoints, the RN recursion can be interpreted as neutral growth after a charged seed has already been formed. The seed affects the lower cutoff, while the large mass logarithmic term is controlled by the fixed charge recursion.

The charged history analysis has been corrected relative to its interpretation. Ordered charged histories produce a Gaussian charge distribution and a history entropy whose leading slope is $\ln 3$ per step. This factor is the entropy of the formal unrestricted three channel alphabet. It is not a proof that the same formula is the microscopic degeneracy of a RN \BH{}.

Two different absorption histories may lead to the same observed coarse grained result after hidden records are traced out. They do not become the same exact quantum state in a unitary description. The missing information is stored in hidden, environmental, or radiation degrees of freedom. Therefore the relevant degeneracy is a coarse grained endpoint degeneracy.

The toy calculations show that minimal history uniqueness is not generic. In a state function model, many ordered sign histories can share the same final configuration. Adding a mass or energy label reduces the degeneracy but does not remove it. If local first law updates are made path dependent, the final mass can depend on order, which further shows that history entropy and state entropy are not automatically equal. A unique minimal history can be obtained only after imposing an extra canonical protocol.

These results preserve the useful part of the original construction while making its assumptions explicit. The model should be understood as an effective description of discrete \BH{} growth. It corrects the asymptotics, separates dynamical entropy from history entropy, and makes the normalization assumptions visible.

\appendix

\section{Asymptotic inversion of the Schwarzschild--Tangherlini recursion}\label{app:general_d}

This appendix gives the technical step behind Eq.~\eqref{eq:N_general_corrected}. Write the discrete recursion in the form
\begin{align}
M_{N+1}&=M_N+f(M_N),\\
f(M)&=K_d M^{-s},
\qquad
s=\frac{1}{d-2}.
\end{align}
Let $N(M)$ denote the asymptotic inverse of the sequence. It satisfies
\begin{equation}
N(M+f(M))-N(M)=1.
\end{equation}
Expanding the left-hand side for large $M$ gives
\begin{equation}
f(M)N'(M)+\frac{f(M)^2}{2}N''(M)+\cdots=1.
\end{equation}
If $N_0'(M)=1/f(M)$ is the continuum contribution, then the next term must cancel
\begin{equation}
\frac{f(M)^2}{2}N_0''(M)=-\frac{f'(M)}{2}.
\end{equation}
Thus the first discrete correction obeys
\begin{align}
f(M)N_1'(M)&=\frac{f'(M)}{2},\\
N_1(M)&=\frac{1}{2}\ln f(M)+\text{const.}.
\end{align}
Since $f(M)=K_dM^{-1/(d-2)}$, this gives
\begin{equation}
N_1(M)=-\frac{1}{2(d-2)}\ln M+\text{const.}.
\end{equation}
Adding the continuum integral then yields Eq.~\eqref{eq:N_general_corrected}.

\section{Fixed charge Reissner--Nordstr\"om asymptotics}\label{app:rn_derivation}

For the RN sector, the step is
\begin{align}
f(M,Q)&=\Delta M(M,Q)=\frac{K_{\mathrm{eff}}}{r_+(M,Q)},\\
r_+(M,Q)&=M+\sqrt{M^2-Q^2}.
\end{align}
At fixed $Q$ and large $M$,
\begin{equation}
r_+(M,Q)=2M-\frac{Q^2}{2M}-\frac{Q^4}{8M^3}+\Order(M^{-5}).
\end{equation}
The continuum part of the inverse recursion is therefore
\begin{align}
\int^M \frac{\dd x}{f(x,Q)}
&=\frac{1}{K_{\mathrm{eff}}}\int^M r_+(x,Q)\,\dd x \\
&=\frac{M^2}{K_{\mathrm{eff}}}-\frac{Q^2}{2K_{\mathrm{eff}}}\ln M\notag\\
&\quad+\Order(M^{-2})+\text{const.}.
\end{align}
The discrete correction found in Appendix~\ref{app:general_d} is again 
\begin{equation}
    \frac12\ln f(M,Q).
\end{equation}
Since
\begin{equation}
f(M,Q)=\frac{K_{\mathrm{eff}}}{2M}\left[1+\Order(M^{-2})\right],
\end{equation}
this contributes
\begin{equation}
\frac{1}{2}\ln f(M,Q)=-\frac{1}{2}\ln M+\Order(M^{-2})+\text{const.}.
\end{equation}
Combining the continuum and discrete logarithms gives
\begin{equation}
B_{\mathrm{RN}}(Q)=-\frac{1}{2}-\frac{Q^2}{2K_{\mathrm{eff}}},
\end{equation}
which is Eq.~\eqref{eq:RN_log_coeff}.

\section{Saddle-point counting of charged histories}\label{app:history_derivation}

For histories with elementary charges $+, -,$ and $0$, define
\begin{align}
p_+&=\frac{N_+}{N},
\quad p_-=\frac{N_-}{N},\\
p_0&=\frac{N_0}{N},
\quad q=\frac{Q}{\ech N}=p_+-p_-.
\end{align}
The leading Stirling approximation to the multinomial entropy is
\begin{equation}
\frac{S_{\mathrm{hist}}}{\kb N}= -p_+\ln p_+-p_-\ln p_- -p_0\ln p_0.
\end{equation}
Maximizing this expression at fixed $q$ gives
\begin{align}
p_+&=\frac{\ee^{\lambda}}{1+2\cosh\lambda},\\
p_-&=\frac{\ee^{-\lambda}}{1+2\cosh\lambda},\\
p_0&=\frac{1}{1+2\cosh\lambda}.
\end{align}
where
\begin{equation}
q=\frac{2\sinh\lambda}{1+2\cosh\lambda}.
\end{equation}
For small $q$, one has $\lambda=3q/2+\Order(q^3)$, and the entropy density becomes
\begin{equation}
\frac{S_{\mathrm{hist}}}{\kb N}=\ln 3-\frac{3}{4}q^2+\Order(q^4).
\end{equation}
Restoring $q=Q/(\ech N)$ gives the leading terms of Eq.~\eqref{eq:Shist_main}. The subleading term $-\frac12\ln N$ comes from the Gaussian fluctuation determinant in the constrained multinomial sum.

\end{document}